\documentclass[twocolumn,pre]{revtex4}
\usepackage{graphicx}
\usepackage{amssymb,amsmath,enumerate}
\usepackage{color}

\newcommand{\be}{\begin{eqnarray}}
\newcommand{\ee}{\end{eqnarray}}
\newcommand{\D}{\mathrm{d}}

\begin{document}

\title{Fundamental challenges in packing problems: from spherical to non-spherical particles}

\author{Adrian Baule$^1$ \& Hern\'an A. Makse$^{2}\footnote{Correspondence to: hmakse@lev.ccny.cuny.edu}$ }

\affiliation{
$^1$School of Mathematical Sciences, Queen Mary University of London, Mile End Road, London E1 4NS, UK\\
$^2$Levich Institute and Physics Department, City College of New York, New York, New York 10031, USA
}

\begin{abstract}

Random packings of objects of a particular shape are ubiquitous in science and engineering. However, such jammed matter states have eluded any systematic theoretical treatment due to the strong positional and orientational correlations involved. In recent years progress on a fundamental description of jammed matter could be made by starting from a constant volume ensemble in the spirit of conventional statistical mechanics. Recent work has shown that this approach, first introduced by S.~F. Edwards more than two decades ago, can be cast into a predictive framework to calculate the packing fractions of both spherical and non-spherical particles.

\end{abstract}

\date{\today}

\maketitle

\section*{Introduction}

In 1989 Edwards and Oakeshott made the remarkable proposal that the macroscopic properties of static granular matter can be calculated as ensemble averages over equiprobable jammed microstates controlled by the system volume \cite{Edwards89}. In other words, granular matter is amenable to a statistical mechanical treatment, where the role of energy is played by the volume. Clearly, there is no a priori reason why such a treatment should be correct. Granular matter is profoundly out of equilibrium, since thermal fluctuations are essentially absent for the macroscopic length scales considered. In particular, there is no equivalent of Liouville's theorem for equilibrium systems due to the strongly dissipative nature of granular assemblies, which are dominated by static frictional forces. Nevertheless, the Edwards' ensemble approach has proven exceedingly useful in characterizing the properties of these athermal states of matter and continues to intrigue both experimentalists and theoreticians alike.

The main statements of this approach are \cite{Edwards89,Edwards94}: (i) The distribution of jammed microstates is flat and independent of the compaction history leading to a natural definition of a configurational entropy $S=\ln \Omega_{\rm Edw}$, where $\Omega_{\rm Edw}$ is the number of jammed configurations. (ii) There is an equivalence between ensemble averages and time averages, if the system can explore its jammed configurations by some external drive (tapping or slow shearing). (iii) The compactivity $X^{-1}=\partial S/\partial V$ characterizes the packing state analogous to the temperature in equilibrium systems. These strong assumptions have been scrutinized in various studies over recent years in order to obtain insight into the validity of Edwards' approach \cite{Ciamarra12}. Soft compaction experiments under continuous tapping have provided evidence for a reversible branch in the packing fraction for a variation of the tapping amplitude, indicating the existence of thermodynamic states \cite{Nowak98,Philippe02,Brujic05,Richard05}. Simple models of such a compaction dynamics have confirmed ergodicity and have been connected to a slow relaxation dynamics akin to the relaxation in glasses \cite{Nicodemi99,Barrat00,Brey00,Dean01,Danna01}. One signature of such a slow dynamics is the existence of a non-equilibrium fluctuation-dissipation relation \cite{Nicodemi99,Barrat00,Makse02}. Indeed, the effective temperature appearing in FDRs under perturbations agrees with the configurational temperature $T^{-1}_{\rm eff}=\partial S/\partial E$ in Edwards' framework \cite{Makse02}.

Ergodicity has also been demonstrated explicitly in more realistic simulations \cite{Ciamarra06,Wang12}. The compactivity has been measured in simulations and experiments \cite{Schroeter05,Lechenault06,Ribiere07,Briscoe08,McNamara09}. On the other hand, results on the equiprobability of microstates are mixed. By evaluating the probabilities of jammed microstates in small clusters a break down of the flat distribution assumption has been demonstrated \cite{Xu05,Gao06,Gao09,Xu11}, which might be traced back to the packing protocol used \cite{Wang12}. Recent studies have investigated the equilibration of granular subsystems in contact \cite{Lechenault10,Wang12,Puckett13}, providing further insight into the thermodynamic nature of granular matter.
 
Ultimately, the success of any statistical mechanical theory needs to be measured by the comparison with experiments. One key problem in Edwards' approach is to identify a suitable {\it volume function}, which parametrizes the total volume of the packing as a function of the particle configurations (positions and orientations), replacing the role of the Hamiltonian \cite{Edwards94}. Here, different conventions can be employed to partition the total volume into cells associated with each particle \cite{Ball02,Blumenfeld03}, the simplest of which is the Voronoi tesselation \cite{Okabe,Makse04}. In 3D these exact volume functions are difficult to handle analytically, so that reduced representations are sought. The thermodynamic nature of packings suggests to use a coarse-grained description of the volume function in terms of observables such as the average number of contacts $z$ (coordination number) \cite{Song08,Song10,Wang11,Baule13}. In turn, $z$ is determined by the force transmission in the contact network leading to a mechanically stable packing in which forces and torques on each particle balance \cite{Alexander98}. For the force network ensemble approaches similar to the volume ensembles have been introduced in order to explain the observed force distribution \cite{Delaney10} from an entropy maximization \cite{Kruyt02,Bagi03,Goddard04,Snoeijer04,Henkes09}. The stress tensor has also been considered as a conserved quantity leading to a different class of ensembles \cite{Henkes07,Henkes09}.

Recently, the force transmission has been treated on a random graph under local mechanical stability constraints resulting in quantitative predictions for the force distribution and the value of $z$ using a cavity method \cite{Bo13}. The problem of finding the densest random packing can be similarly formulated as a constraint optimization problem: Random close packings appear as the groundstate of the volume ensemble restricted to disordered packings as $X\to 0$ for a given $z$ \cite{Song08,Song10}. This picture highlights that jamming falls into the class of NP optimization problems \cite{Krzakala07}, which can be tackled successfully with the methods of statistical mechanics such as cavity methods \cite{Bo13}. A full solution needs to combine the two approaches for the force and volume ensembles, where the Hamiltonian that enforces jamming is a function of both the particle configurations and the contact forces on a random contact network. These ensemble approaches are thus similar in spirit to other recent works that consider jamming as the infinite pressure limit of metastable glass phases \cite{Aste04,Parisi05,Kamien07,Krzakala07,Mari09,Biazzo09,Hermes10,Parisi10}. Here, one considers instead of the Edwards entropy $S$, the ``glass complexity" in order to obtain the statistics of the metastable basins as the pressure diverges. Treatments of this problem based on the random-first order transition picture and replica theory have been performed \cite{Parisi10}.

\section*{From spheres to non-spheres}

Random packings of hard objects appear in a broad range of scientific and engineering fields like self-assembly of nanoparticles, liquid crystals, glass formation and bio-materials \cite{Torquato10}. In fact, the question of how densely objects of a particular shape can fill a given volume is probably one of the most ancient scientific problems and still of great practical importance for all industries involved in granular processing. The densest random packing has been extensively studied in experiments and simulations for spheres, which typically reach a maximal volume fraction of $\phi\approx 0.64$ in monodisperse assemblies \cite{Bernal60,Skoge06}. This value is quite robustly reproducible and commonly referred to as random close packing (RCP) density. However, much less is known about anisotropic shapes, despite the fact that all shapes in nature deviate from the ideal sphere. A theoretical investigation of the packing problem has proven notoriously difficult due to the strong positional and orientational correlations of dense packings. In fact, a mathematically rigorous treatment of random sphere packings has been the outstanding component in T.C.~Hales' proof of Kepler's conjecture on the densest packing of spheres \cite{Hales05}.

Recent empirical work has focused on packings of anisotropic shapes like ellipsoids, spherocylinders, and tetrahedra, which can achieve considerably denser volume fractions than the spherical RCP \cite{Williams03,Abreu03,Donev04,Man05,Jia07,Bargiel08,Wouterse09,HajiAkbari09,Jaoshvili10,Lu10,Kyrylyuk11,Jiao11,Zhao12} (see Table~\ref{table1}). In fact, a conjecture attributed to Ulam in the context of regular packings \cite{Gardner} and recently also formulated for random packings \cite{Jiao11} states that the sphere is, indeed, the worst packing object among all convex shapes. This suggests to improve packing fractions by searching in the space of object shapes, but in the absence of any general theory, this search could so far only be performed on a case-by-case basis using experiments and simulations. A caveat of such empirical studies is the strong protocol dependence of the final close packed state even for the same shape. While the range of achieved volume fractions is relatively small for spheres \cite{Skoge06}, recent studies of spherocylinder packings, e.g., exhibit a much greater variance depending on the algorithm used \cite{Williams03,Abreu03,Jia07,Bargiel08,Wouterse09,Lu10,Kyrylyuk11,Jiao11,Zhao12}. Further theoretical insight is needed, which can be obtained by considering a coarse-grained distribution for the Voronoi volumes in the packing, as discussed next.

\section*{A mean-field theory for random close packings}

In the Voronoi convention one associates with each particle the
fraction of space that is closer to this particle than to any other
one. This defines the Voronoi volume $W_i$ of a particle $i$, which
depends on the configurations of all remaining particles $\mathbf{x}_j=(\mathbf{r}_j,\mathbf{\hat{t}}_j)$,
(including position $\mathbf{r}_j$ and orientation $\mathbf{\hat{t}}_j$).  The total
volume $V$ occupied by $N$ particles is $ V=\sum_{i=1}^N
W_i(\{\mathbf{x}_1,...,\mathbf{x}_N\})$, and the packing fraction of
monodisperse particles of volume $V_\alpha$ is $\phi=N V_\alpha/V$. In order to
determine $W_i$ one has to know the Voronoi boundary (VB) between
two particles $i$ and $j$, which is the hypersurface that contains all
points equidistant to the surfaces of both particles and thus depends on the particle
shape and their relative configuration. The boundary of $W_i$ then follows from a global minimization procedure over all pairwise VB \cite{Song10}. In order to take into account multi-particle correlations in the packing, we use a statistical treatment where the overall volume is expressed in terms of an average Voronoi volume:
$V=N\overline{W}(z)$, so that $\phi=V_\alpha/\overline{W}(z)$. Instead of an exact description in
terms of all configurations $\{\mathbf{x}_1,...,\mathbf{x}_N\}$, the average Voronoi
volume is characterized by the coordination number $z$, which denotes
the average number of contacting neighbours in the packing. We derive a self-consistent equation for the coarse-grained volume function $\overline{W}(z)$ of monodisperse particles \cite{Song08,Song10,Baule13}:
\be
\label{integral}
\overline{W}(z)=\int \D \mathbf{c}\exp\left\{-\frac{V^*(\mathbf{c})}{\overline{W}(z)-V_\alpha}-\sigma(z)\,S^*(\mathbf{c})\right\}.
\ee
Here, $V_\alpha$ is the volume of a single particle and $\sigma(z)$ is the average free-surface of particles at contact, which can be estimated from local configurations of $z$ contacting particles. Formally, the integrand on the right hand side can be considered as the cumulative distribution function $P(\mathbf{c})$ containing the probability to find the boundary of the Voronoi volume in the direction $\mathbf{\hat{c}}$ at a value larger than $c$. This quantity can be interpreted geometrically as the probability to find all $N-1$ particles outside a volume $\Omega$ centred at $\mathbf{c}$ from the reference particle (see Fig.~\ref{Fig_exv}a). The particular form of $P(\mathbf{c})$ results from a factorization into bulk and contact terms, which are motivated from the dominant contributions in the radial distribution function \cite{Song08,Song10,Jin10b,Baule13}.

The quantities $V^*$ and $S^*$ are the Voronoi excluded volume and surface, which extend the usual hard-core excluded volume of equilibrium systems $V_{\rm ex}$ \cite{Onsager49} to packings. The volume $V^*$ is the volume excluded by $\Omega$ for bulk particles and takes into account the overlap between $\Omega$ and $V_{\rm ex}$: $V^*=\overline{\Omega}-\overline{\Omega\cap V_{\rm ex}}$, where the bar denotes an orientational average. Likewise, $S^*$ denotes the surface excluded by $\Omega$ for contacting particles: $S^*=\overline{\partial V_{\rm ex}\cap\Omega}$. Plots of $V^*$ and $S^*$ for spherocylinders are shown in Fig.~\ref{Fig_exv}b. Analytical expressions for $V^*$ and $S^*$ can be derived in the spherical limit in closed form \cite{Song08,Song10}. For non-spherical shapes analytic expressions for the VB can be derived using a suitable decomposition of the shape into overlapping and/or intersecting spheres. This leads to exact expressions for $V^*$ and $S^*$, which can be evaluated numerically \cite{Portal13}. Interestingly, in the limit $\alpha\to 1$, Eq.~(\ref{integral}) admits an exact solution for spheres: $W(z)=2\sqrt{3} V_1/z$. As a consequence, we obtain an equation of state for spherical packings \cite{Song08,Song10}
\be
\label{sphericalrb}
\phi(z)=\frac{z}{z+2\sqrt{3}},
\ee
which predicts the limiting values $\phi=0.536$ and $\phi=0.634$ under the isostatic conditions $z=4$ and $z=6$ for infinitely rough and frictionless spheres, respectively. Using the thermodynamic framework one can show that these two values are reached in the limits of infinite and zero compactivity, respectively \cite{Song08}. Therefore, the spherical equation of state leads to a statistical interpretation of RCP as the ground state of disordered sphere packings. The predictions for the limiting values are in good agreement with the values found in experiments and simulations for both random loose packings and RCP of spheres.

Under deformation from the sphere, higher packing fractions are typically reached, where the spherical limit appears as a singular point in the $\phi(\alpha)$ plane. Moreover, smooth shapes close to the sphere are not isostatic but hypostatic with $z<2d_{\rm f}$ due to redundancies in the force and torque balance equations \cite{Chaikin06,Donev07}. The variation $z(\alpha)$ is obtained by considering the average effective number of degrees of freedom $\tilde{d}_{\rm f}$ defined as the number of linearly independent force and torque balance equations: $z=2\left<\tilde{d}_{\rm f}(\alpha)\right>$ \cite{Baule13}. Here, the probability of redundant configurations with $\tilde{d}_{\rm f}<d_{\rm f}$ can be estimated by re-weighting all configurations by rotating into states of maximal redundancy.

The existence of redundant configurations explains the observed convergence in $z(\alpha)$ to values close to $8$ for spherocylinders with large aspect ratios \cite{Wouterse09,Zhao12}: For long spherocylinders the contacts are predominantly on the cylindrical part so that all normal forces are coplanar. As a consequence, the effective number of degrees of freedom is reduced by one, leading to $z=8$ \cite{Baule13}. The requirement of local force and torque balance can also be formulated as a constraint optimization problem on a factor graph, which describes the force transmission on a single particle \cite{Bo13}. Solving this problem with standard methods such as the cavity method predicts values of $z$ in frictional packings and also allows for the computation of the distribution of contact forces in good agreement with experimental results \cite{Bo13}.

\section*{Phase-diagram of jammed isotropic and anisotropic particles}

The combination of the results for $\overline{W}(z)$ and $z(\alpha)$ leads to a complete theoretical prediction for the packing density $\phi(\alpha)=V_\alpha/\overline{W}(z(\alpha))$ of non-spherical particles without any adjustable parameters \cite{Baule13}. We estimate the maximum density of spherocylinders at $\alpha = 1.3$ with a density $\phi_{\rm max} = 0.731$ in good agreement with empirical data. The theory also reproduces well the density of dimers, estimating a maximum at $\alpha=1.3$ with $\phi_{\rm max} \approx 0.707$. We have also calculated the packing fraction of lens-shaped particles, which can serve as approximations for oblate ellipsoidal shapes. Our theory yields $\phi_{\rm max}=0.736$ for $\alpha=0.8$. This shape represents the densest random packing of an axisymmetric shape known so far. The appearance of a maximum in $\phi$ for non-spherical shapes close to the sphere has been explained in a simple qualitative picture on the basis of the excluded volume $V_{\rm ex}$ \cite{Williams03}. For $\alpha$ close to $1$, the ratio $V_{\rm ex}/V_\alpha$ changes only slightly from the spherical value and a density increase results due to the additional orientational degrees of freedom, whereby the particles can fit into gaps by rotating, similar to the increase in packing efficiency due to polydispersity \cite{Danisch10}. For larger $\alpha$, $V_{\rm ex}$ exceeds $V_\alpha$ while $z$ remains constant, so that the packing is dominated by the excluded volume and the packing fraction decreases. This argument can explain qualitatively the observed larger packing fraction of spherocylinders compared with dimers. The ratio $V_{\rm ex}/V_\alpha$ is approximately equal for both shapes up to $\alpha\approx1.2$, but for larger $\alpha$ the ratio for dimers increases beyond that of spherocylinders. The packing densities derived in our framework are interpreted as upper bounds to the empirically obtained densities and correspond to maximally random jammed states \cite{Torquato00} by construction, since the distribution of contact angles in the first coordination shell is imposed to be uniform, avoiding any partial order.

By plotting $z(\alpha)$ against $\phi(\alpha)$ parametrically as a function of $\alpha$, we obtain a phase diagram in the $z$-$\phi$ plane (Fig.~\ref{Fig_pd}). Surprisingly, we find that both dimer and spherocylinder packings appear as smooth continuations of spherical packings. The analytic form of this continuation from the spherical random branch can be derived (blue dashed line in Fig.~\ref{Fig_pd}) \cite{Baule13}. A comparison of our theoretical results with empirical data for a large variety of shapes highlights that the analytic continuation provides a boundary line in the $z$-$\phi$ phase diagram. Maximally dense disordered packings appear to the left of this boundary, while the  packings to the right of it are partially ordered. The spherical ordered branch provides another boundary, which separates tetrahedra from all other shapes: Tetrahedra are the only shape that pack in a disordered way denser than spheres in a FCC crystal. We observe that the maximally dense packings of dimers, spherocylinders, lens-shaped particles and tetrahedra all lie surprisingly close to the analytic continuation of RCP. Whether there is any deeper meaning to this remains an open question.

The picture that emerges is that spherical packings can be generated between the RLP and RCP limits by variation of the inter-particle friction, since this leads to an increase in the coordination number under the isostatic condition from $z=4$ to $z=6$. Beyond RCP, the spherical equation of state can be continued smoothly by deforming the sphere into elongated shapes. Moreover, the spherical RCP is interpreted as the freezing point of disordered sphere packings, associated with a melting point at $\phi=0.68$ \cite{Radin08,Jin10}. The signature of this disorder-order transition is a discontinuity in the entropy density of jammed configurations as a function of the compactivity. This highlights the fact that beyond RCP, denser packing fractions of spheres can only be reached by partial crystallization up to the homogeneous FCC crystal phase \cite{Torquato00}.

\section*{Conclusions and outlook}

The first-order transition of jammed spheres identified within Edwards' thermodynamics \cite{Jin10} is reminiscent of the entropy induced phase transition of equilibrium hard spheres, which is found at $\phi=0.494$ and $\phi=0.545$, respectively. However, it should be emphasized that the physical origin of these two transitions is fundamentally different: The equilibrium phase transition is a consequence of the maximization of the conventional entropy, while the transition at RCP of jammed spheres is driven by the competition between volume minimization and maximization of the entropy of jammed configurations $S$. For anisotropic particles at equilibrium, a disorder-order phase transition appears, e.g., between isotropic and nematic phases of elongated shapes: For large $\alpha$, Onsager's theory of equilibrium hard rods predicts a first order isotropic-nematic transition with freezing point at the rescaled density $\phi \alpha=3.29$ and melting point at $\phi \alpha=4.19$ \cite{Onsager49}. For colloidal suspensions of more complex shapes like polyhedra, both liquid crystal as well as plastic crystal and even quasicrystal phases have been found \cite{Escobedo11,Damasceno12,Marechal13}. By analogy with the case of jammed spheres, one might wonder whether packings of non-spherical particles exhibit similar transitions that might be characterized in the $z$-$\phi$ phase diagram. Packings of hard thin rods indeed satisfy a scaling law, where the RCP has been experimentally identified at $\phi\alpha \approx 5.4$ \cite{Philipse96}. 

The Edwards' approach thus helps to elucidate how macroscopic properties of granular matter arise from the anisotropy of the constituents -- one of the central questions in present day materials science \cite{Glotzer07,Borzsonyi13}. A better understanding of this problem will facilitate, e.g., the engineering of new functional materials with particular mechanical responses by tuning the shape of the building blocks. A search in the space of object shapes for optimization can be performed by considering a small number of spheres and systematically exploring the different possible configurations \cite{Miskin13}. 

\begin{figure*}
\centering
\includegraphics[width=11cm]{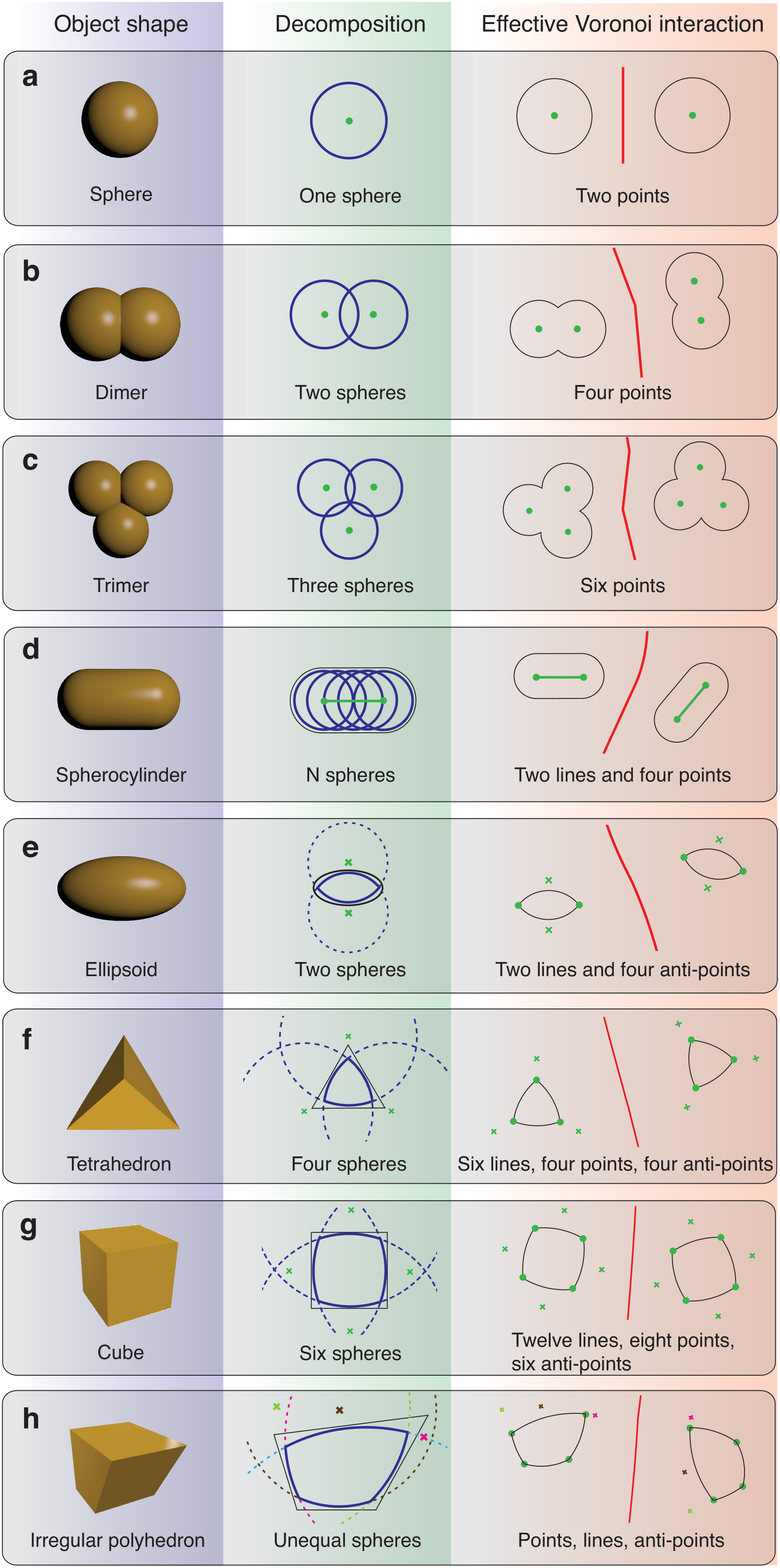}
\caption{\label{Fig_shapes}}
\end{figure*}

Our approach Eq.~(\ref{integral}) can be applied to a large variety of both convex and non-convex shapes. The key is to parametrize the Voronoi boundary between two such shapes, which allows for the calculation of the Voronoi excluded volume and surface. In fact, analytical expressions for the Voronoi boundary can be derived following an exact algorithm for arbitrary shapes by decomposing the shape into overlapping and intersecting spheres (see Fig.~\ref{Fig_shapes}). Therefore, a systematic search for maximally dense packings in the space of given object shapes can be performed using our framework. Extensions to mixtures and polydisperse packings can also be formulated. So far, exhaustive searches for dense packings have only been performed for ordered packings using computer simulations \cite{Graaf11,Graaf12} and a combination of analytic and simulation techniques \cite{Chen13}. This has elucidated in particular the validity of Ulam's conjecture that the sphere is the worst packing object in 3d \cite{Gardner}. Analytical progress to prove this conjecture locally, that is, for shapes deformed from the sphere, has recently been made \cite{Kallus13}.

A more systematic investigation of disordered packings can shed light on the validity of a random variant of Ulam's conjecture, which so far has only been investigated in simulations \cite{Jiao11}. Our analytic continuation from RCP highlights that this conjecture might hold more generally than previously assumed, containing not only convex shapes, but also a significant class of non-convex ones. Ultimately, our approach might lead to a more exhaustive theoretical investigation of Ulam's conjecture. Along the way one might be able to answer important questions such as if a shape that packs denser in a random configuration than in a regular one exists \cite{Chaikin06}. Such objects could represent optimal glass formers with far reaching consequences for materials science.


\clearpage

TABLE~\ref{table1}: {\bf Overview of maximal packing fractions found for disordered packings of a selection of regular shapes.} We observe that spheres are the worst-packing objects among all shapes in 3d as conjectured by Ulam \cite{Gardner}, while tetrahedra achieve the densest disordered packing. We note that the tetrahedron is the only shape known that packs in a disordered arrangement denser than spheres in the FCC crystal ($\phi_{\rm FCC}=0.7405$). Ellipsoids and lens-shaped particles pack very close to this value.

\begin{figure*}
\centering
\includegraphics[width=15cm]{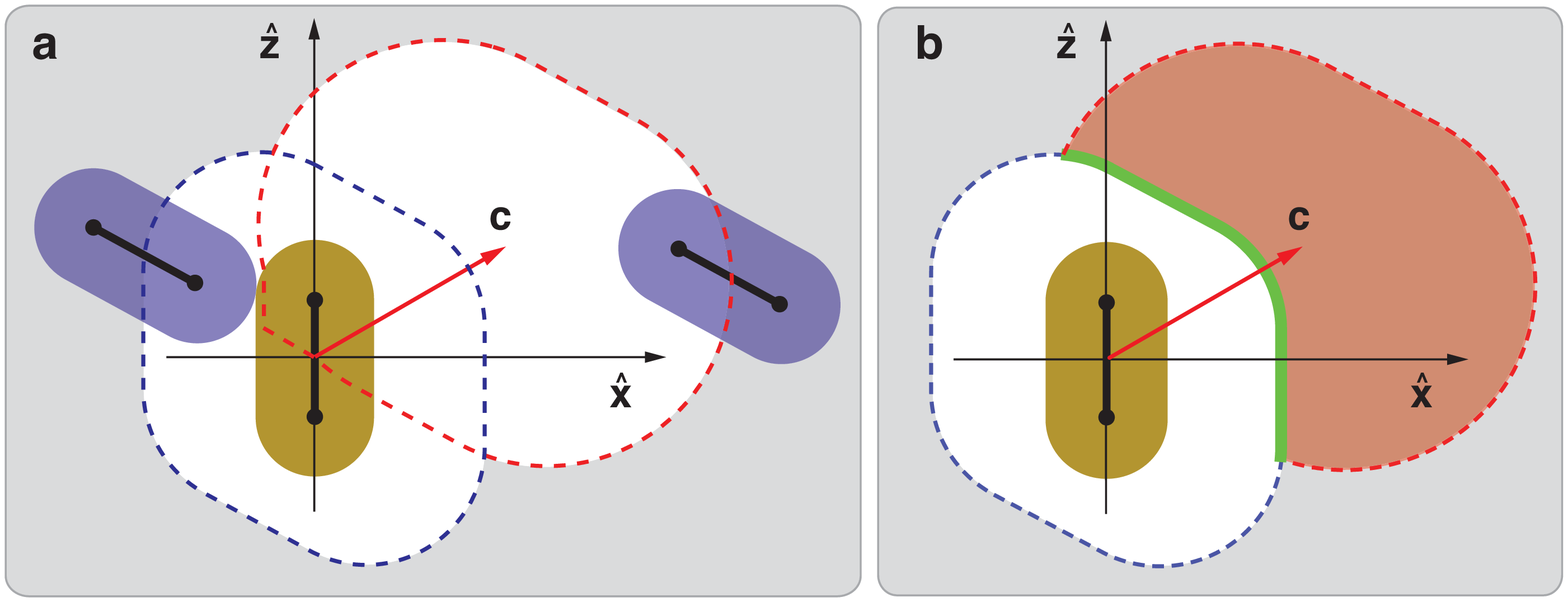}
\caption{\label{Fig_exv}}
\end{figure*}

FIG.~\ref{Fig_exv}: {\bf The Voronoi excluded volume and surface for spherocylinders.} {\bf (a)} The volume $\Omega$ (red) is excluded for the remaining $N-1$ particles in the packing because otherwise the Voronoi boundary would be found at a value smaller than $c$ in the direction $\mathbf{\hat{c}}$. We draw the usual hard-core excluded volume $V_{\rm ex}$ \cite{Onsager49} in blue. {\bf (b)} The overlap of $\Omega$ and $V_{\rm ex}$ defines the Voronoi excluded volume $V^*$ (red) and the Voronoi excluded surface $S^*$ (green). Figure taken from Ref.~\cite{Baule13}.

\begin{figure*}
\centering
\includegraphics[width=16cm]{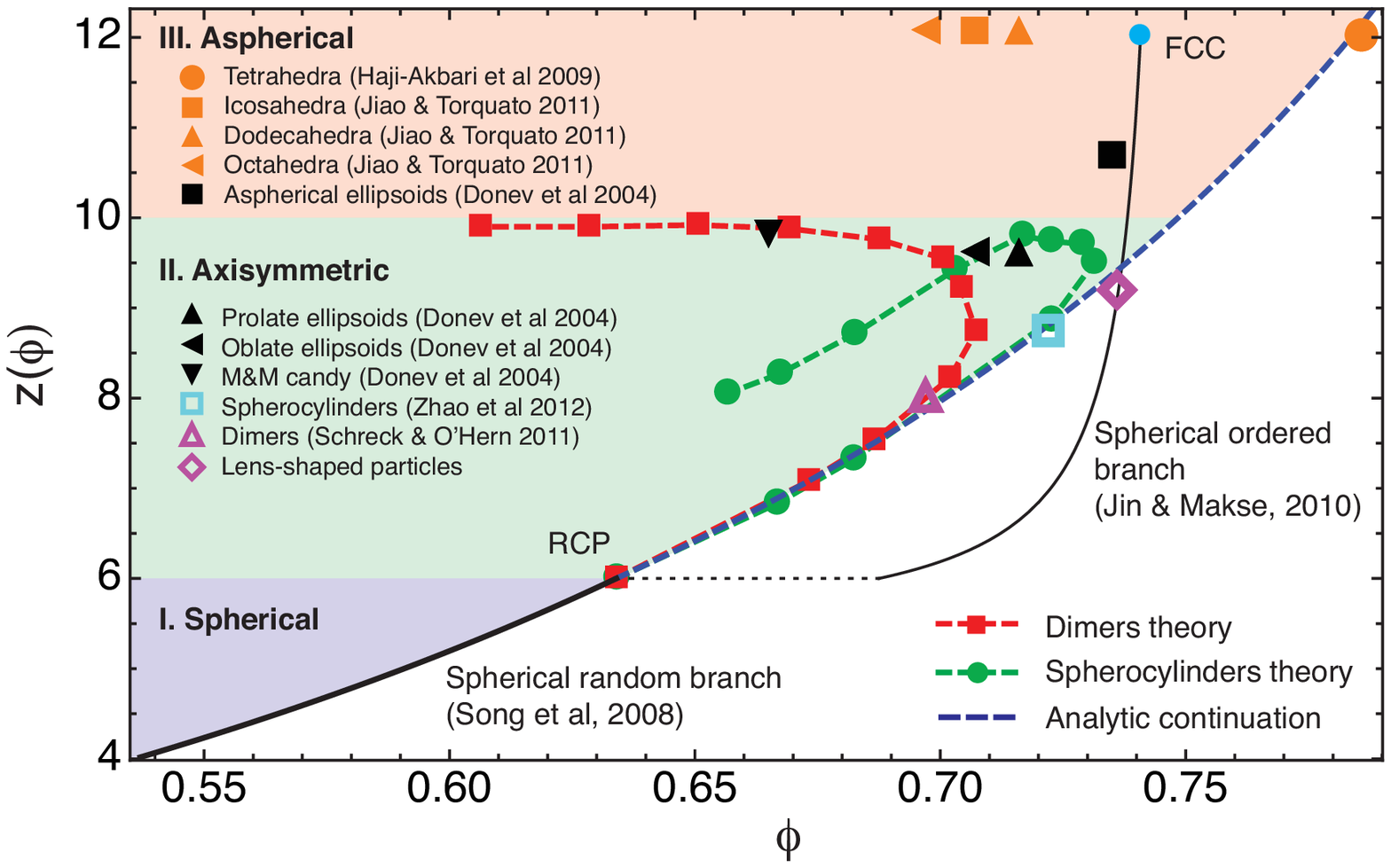}
\caption{\label{Fig_pd}}
\end{figure*}

FIG.~\ref{Fig_pd}: {\bf Phase diagram of jammed matter.} We plot our results for dimers and spherocylinders in the $z$-$\phi$ plane together with results from the literature for frictionless disordered packings of a selection of regular shapes. We have selected those shapes for which the $z$ and $\phi$ values have been determined in the same simulation. The predicted spherical random branch Eq.~(\ref{sphericalrb}) \cite{Song08} (thick black line )and a conjectured first order disorder--order transition at RCP for spheres \cite{Jin10} (dotted and thin black lines) are also indicated. We observe that the analytic continuation of RCP under deformation into dimers and spherocylinders provides an empirical bound to disordered packings in the phase diagram. The symmetry of the shape indicates the possible values of the coordination number $z$: (i) Spheres have $z$ between $4$ (infinitely rough) and $6$ (frictionless). (ii) Axisymmetric particles have $z$ between $6$ and $10$. (iii) Fully aspherical particles have $z$ between $10$ and $12$. Note that for polyhedra, $z$ is associated with the total degrees of freedom blocked by the different types of contacts (face-face, face-vertex, vertex-vertex, face-edge) \cite{Jiao11}. The data point for lens-shaped particles is a theoretical prediction \cite{Baule13}.

FIG.~\ref{Fig_shapes}: {\bf Decomposition of various shapes to calculate the Voronoi boundary.} The Voronoi boundary (VB) between two particles is defined as the hypersurface that contains all points equidistant to their surfaces. This implies that the VB between two equal spheres, e.g., is that between two points at the centers of the spheres, so that the VB is generated effectively by the interaction of two points (a). Likewise, the VB between two spherocylinders is due to the effective interaction of two lines, since spherocylinders can be represented as dense overlaps of spheres (d). Arbitrary shapes can be decomposed into dense overlaps of spheres following certain design principles \cite{Phillips12}. The VB between two such shapes can then be calculated following an exact algorithm that considers the effective Voronoi interactions between points and lines (a -- d) \cite{Baule13}. For shapes that are not naturally given as overlapping spheres (e -- h), we propose alternatively an approximation in terms of a small number of intersecting spheres. In this way, two intersecting spheres (a lens-shaped particle) approximate an oblate ellipsoid and four intersecting spheres approximate a tetrahedron. The effective Voronoi interactions are then between points, lines, and anti-points (indicated by crosses) \cite{Baule13}. Anti-points arise from the inversion of the effective interaction between the spheres in the decomposition. This is evident in the case of lens-shaped particles (e), where the interaction between the spheres is inverted compared to the case of dimers (b). The VB between two tetrahedra is then due to the interaction between the vertices (leading to four point interactions), the edges (leading to six line interactions), and the faces (leading to four anti-point interactions). This approach can be generalized to arbitrary polyhedra. Figure taken from Ref.~\cite{Baule13}.

\begin{table*}[ht]
\centering
\begin{tabular}{c | c | c | c}
Shape  & $\phi_{\rm max}$ simulation & $\phi_{\rm max}$ experiment & $\phi_{\rm max}$ theory \\ 
\hline \hline
sphere &  0.645 \cite{Skoge06} & 0.64 \cite{Bernal60} & 0.634 \cite{Song08} \\
\hline
M\&M candy & & 0.665 \cite{Donev04} & \\
dimer & 0.703 \cite{Faure09} &  &  0.707 \cite{Baule13} \\
oblate ellipsoid & 0.707 \cite{Donev04} &  &  \\
prolate ellipsoid & 0.716 \cite{Donev04}  & &   \\
spherocylinder & 0.722 \cite{Zhao12} &  & 0.731 \cite{Baule13}  \\
lens-shaped particle & & & 0.736 \cite{Baule13}   \\
\hline
octahedron & 0.697 \cite{Jiao11} & & \\
icosahedron & 0.707 \cite{Jiao11} &  & \\
dodecahedron & 0.716 \cite{Jiao11} & & \\
general ellipsoid & 0.735 \cite{Donev04} & 0.74 \cite{Man05} &   \\
tetrahedron & 0.7858 \cite{HajiAkbari09} & 0.76 \cite{Jaoshvili10} &
\end{tabular}
\caption{\label{table1}}
\end{table*}

\end{document}